\documentclass[12pt,a4paper]{article}
\usepackage{amssymb,graphicx,graphics,epsfig}
\usepackage{lscape,graphics,amsmath}
\usepackage{latexsym,amsfonts}
\usepackage{cite}
\usepackage{textcomp}
\usepackage{float}
\begin{document}
\title{\vspace{-1cm}\bf Quantization of electromagnetic field\\ in the Schwarzschild spacetime}

\author{
Vadim Egorov, Mikhail Smolyakov, Igor Volobuev
\\
{\small{\em Skobeltsyn Institute of Nuclear Physics, Lomonosov Moscow
State University,
}}\\
{\small{\em Moscow 119991, Russia}}}

\date{}
\maketitle
\begin{abstract}
We discuss the problem of canonical quantization of
electromagnetic field in the Schwarzschild spacetime. It is shown
that a consistent procedure  of canonical quantization of the
field can be carried out without taking into account the internal
region of the black hole. We prove that there exists a unitary
gauge, which can be viewed as a combination of the Coulomb and
Poincare gauges and is compatible with the field equations. The
solutions corresponding to the stationary one-particle states of
the electromagnetic field are studied, the canonical commutation
relations and the Hamiltonian of the quantized electromagnetic
field are obtained.
\end{abstract}

\section{Introduction}
Starting from the well-known papers
\cite{Boulware:1974dm,Hartle:1976tp}, the problem of field
quantization in the presence of black holes is widely discussed in
the literature. However, many questions still remain
insufficiently clarified. In particular, there exist conceptual
problems like the existence of the so-called ``white hole'' in the
Kruskal-Szekeres coordinates \cite{Kruskal:1959vx,Szekeres:1960gm}
and a well-known problem with locality associated with the
location of the white hole in our Universe or even in a parallel
world. One of the attempts to solve this problem
\cite{tHooft:2018waj,tHooft:2019xwm,Hooft:2022azz} led to the idea
of ``quantum cloning'' of the outer regions of black and white
holes. In this approach, the inner regions of black and white
holes turn out to be only mathematical artifacts that have no
direct physical interpretation and do not play any role in the
evolution of the system \cite{Hooft:2022azz}.

The conclusion of paper \cite{Hooft:2022azz}, together with the
approach used in recent papers
\cite{Akhmedov:2020ryq,Anempodistov:2020oki,Bazarov:2021rrb},
motivated us to consider quantum field theory only in the region
outside the horizon of the Schwarzschild black hole. The idea was
to carry out a mathematically rigorous procedure of canonical
quantization, including checking the fulfillment of all the
necessary commutation relations and obtaining the Hamiltonian.
Although the properties of the spectrum of states within the
framework of relativistic quantum mechanics in the gravitational
field of the Schwarzschild black hole are studied quite well, see
\cite{Zecca:2009zz,Barranco:2011eyw}, in general such a treatment
lacks in the literature. For example, in classical paper
\cite{Boulware:1974dm} the fulfillment of the commutation
relations is not checked.

A mathematically rigorous procedure of canonical quantization of a
free real massive scalar field was carried out in papers
\cite{Egorov:2022hgg,Smolyakov:2023pml}. It was shown that in the
resulting quantum theory the canonical commutation relations are
satisfied exactly and the Hamiltonian has the standard form
without any peculiarities. However, an additional twofold
degeneracy of quantum states in comparison with the case of
Minkowski spacetime was found. For the massive spinor field, an
analogous procedure was carried out  in paper
\cite{Egorov:2023ruz}. Based on the results presented in papers
\cite{Egorov:2022hgg,Smolyakov:2023pml,Egorov:2023ruz}, in the
present paper we consider the canonical quantization of
electromagnetic field in the Schwarzschild spacetime. It should be
noted that canonical quantization of electromagnetic field above
the horizon of the Schwarzschild black hole was discussed in the
literature earlier, see \cite{Crispino:1998hp,Crispino:2000jx}. In
these papers, the corresponding equations of motion for
electromagnetic field in the Feynman gauge were solved and the
canonical commutation relations were calculated. In the present
paper, we take the standard action without a gauge fixing term,
leading to the standard ``Maxwell'' form of equations of motion
(like in \cite{Cognola:1997xp}), and consider a different gauge.
We focus on a mathematically rigorous procedure of canonical
quantization, i.e., we perform all the steps which are necessary
for a consistent quantization of the field. Namely, we find a
useful gauge that is compatible with the equations of motion and
prove that this gauge can be imposed, decouple and solve the
corresponding equations of motion, isolate the physical degrees of
freedom and calculate the resulting Hamiltonian, as well as the
canonical commutation relations for them. Thus, it will be
shown below that, like in the cases of scalar and spinor fields, a
consistent quantum field theory can be constructed in the case of
electromagnetic field too, again leading to the doubling of
quantum states in comparison with the case of Minkowski spacetime.

\section{Setup}
The standard metric of the Schwarzschild spacetime in
Schwarzschild coordinates looks like
\begin{equation}\label{metric_Sch}
ds^2=\left( {1 - \frac{{r_0 }}{r}} \right){d}t^2  - \frac{dr^2}{1
- \frac{r_0}{r}} - r^2 \left( {{d}\theta^2  + \sin^2\theta
\,{d}\varphi ^2 } \right),
\end{equation}
where $r_0$ is the Schwarzschild radius. We restrict ourselves to
the region $r > r_0$.

The action of the electromagnetic field in arbitrary curvilinear
coordinates has the form
\begin{equation} \label{em_action}
S=-\frac{1}{4}\int{F_{\mu\nu}F^{\mu\nu}\sqrt{-g}\,{d}^4 x},
\end{equation}
where $F_{\mu \nu}=\partial_\mu A_\nu-\partial_\nu A_\mu$. By
varying action \eqref{em_action} with respect to the vector field
and taking into account that the surface terms at the event
horizon and infinity do not contribute, as well as the
Schwarzschild metric being Ricci-flat, we obtain the equations of
motion in the form
\begin{equation} \label{EoM}
\nabla ^\mu  F_{\mu \nu }  = \nabla ^\mu  \nabla _\mu  A_\nu   -
\partial _\nu  \nabla ^\mu  A_\mu   = 0,
\end{equation}
where $\nabla _\mu $ is the covariant derivative, and the Greek
indices take values $t,r,\theta,\varphi$ for the Schwarzschild
coordinates.

\section{Gauge condition and equations of motion}
In order to proceed further, we need to impose a gauge condition.
Equation \eqref{EoM} suggests that we can impose the Lorentz gauge
condition. However, it seems to be inconvenient in the case under
consideration, because we have to solve the wave equations for
four vector-potential components, which may have different
solutions (similar to equations \eqref{EoM_r}--\eqref{EoM_phi}
below). Then the gauge condition for the quantized field would
include not only the creation and annihilation operators of the
temporal and longitudinal photons but also the wave functions of
these photons, which would complicate the construction of the
physical states (see, for example, \cite{Crispino:2000jx}). For
this reason we prefer to impose a unitary gauge, where such a
problem does not arise.

The gauge transformations of the field $A_{\mu}$ have the standard
form
\begin{equation}
A'_{\mu}(t,r,\theta,\varphi)=A_{\mu}(t,r,\theta,\varphi)-\partial_{\mu}\alpha(t,r,\theta,\varphi).
\end{equation}
First we perform a gauge transformation to put $A_t = 0$ (Weyl
gauge), which is always possible. After imposing this gauge, we
are left with the gauge transformations containing  arbitrary
time-independent gauge functions $\alpha(r,\theta,\varphi)$.

Now let us consider equation \eqref{EoM} with $A_t = 0$ for $\nu =
t$, which reads
\begin{equation} \label{EoM_t}
\frac{{r_0 }}{{r^2 }}\,\partial _t A_r  +
\partial_t\left(\nabla^{\mu}A_{\mu}\right)=0
\end{equation}
(note that $\nabla^{t}A_{t}\neq 0$ for $A_t=0$). From equation
\eqref{EoM_t} it follows that
\begin{equation}\label{pregauge}
\nabla^{\mu}A_{\mu}+\frac{r_0}{r^2}A_r=f\left(r,\theta,\varphi\right),
\end{equation}
i.e., that $\nabla^{\mu}A_{\mu}+\frac{r_0}{r^2}A_r$ does not
depend on time $t$. Formula \eqref{pregauge} suggests the gauge
condition
\begin{equation} \label{gauge2}
\nabla^{\mu}A_{\mu}+\frac{r_0}{r^2}A_r=0.
\end{equation}
One can impose this gauge condition preserving the condition $A_t
= 0$ (see detailed calculations in
Appendix~A).\footnote{Condition \eqref{pregauge} was also
derived from equations of motion in paper \cite{Cognola:1997xp}.
However, in fact the function $f\left(r,\theta,\varphi\right)$ was
set to zero in \cite{Cognola:1997xp} without a proof that this can
actually be done.}

It is worth explaining the physical meaning of gauge condition
\eqref{gauge2}. To this end, let us pass from the Schwarzschild
coordinates $\{ t, r, \theta, \varphi \}$ to the isotropic
coordinates $\{t,X^{1},X^{2},X^{3}\}$, which is carried out
according to the formulas
\begin{equation}\label{Rrtransform}
r = R\left(1+\frac{r_0}{4R}\right)^2
\end{equation}
and
\begin{equation}\label{RtransformXYZ}
X^{1}=R\sin\theta\cos\varphi,\quad
X^{2}=R\sin\theta\sin\varphi,\quad X^{3}=R\cos\theta.
\end{equation}
Then the explicit form of gauge condition \eqref{gauge2} in the
isotropic coordinates can be written as
\begin{equation}\label{gaugeisotr}
{\rm div}\vec A+\frac{r_{0}\left(\frac{r_{0}}{4R}-2\right)}{2R^{3}\left(1-\left(\frac{r_{0}}{4R}\right)^{2}\right)}(\vec R \vec A)=0,
\end{equation}
where $\vec R=(X^{1},X^{2},X^{3})$ and $\vec A=(A_1,A_2,A_3)$
(see detailed calculations in Appendix~B). One can see
that at spatial infinity ($R\to\infty$) this gauge tends to the
Coulomb gauge ${\rm div}\vec A = 0$, while at the horizon ($R\to
r_0/4$) it tends to the Poincare gauge $(\vec R \vec A)=0$. Thus,
gauge condition \eqref{gauge2} is a combination of the Coulomb and
Poincare gauges.

Now we return to the Schwarzschild coordinates. The explicit form
of gauge condition \eqref{gauge2} reads
\begin{equation} \label{gauge_Sch}
\frac{r-r_0}{r^{3}}\,\partial_r\left(r^{2}A_r\right) +
\frac{1}{r^2\sin\theta}\,\partial_{\theta}\left(\sin\theta
A_{\theta}\right)+\frac{1}{{r^2\sin^2 \theta}}\,\partial_\varphi
A_\varphi= 0 .
\end{equation}
Equations \eqref{EoM} for $\nu = r, \theta, \varphi$ are as
follows:
\begin{align}\label{EoM_r}
&\frac{r^{3}}{r-r_{0}}\,\partial _t^2 A_r  -
\frac{1}{\sin\theta}\,\partial_{\theta}\Bigl(\sin\theta\left(\partial_{\theta}A_r-\partial_r
A_{\theta}\right)\Bigr) - \frac{1}{{\sin^2\theta }}\left(
{\partial _\varphi ^2 A_r  - \partial _r \partial _\varphi
A_\varphi  } \right)= 0,
\\\label{EoM_theta}
&\frac{r}{r-r_{0}}\,\partial _t^2
A_\theta-\partial_{r}\left(\frac{r-r_0}{r}\left(\partial_r
A_\theta-\partial_\theta A_r\right)\right) - \frac{1}{{r^2 \sin ^2
\theta }}\left( {\partial _\varphi ^2 A_\theta   - \partial
_\theta  \partial _\varphi  A_\varphi  } \right)=0,
\\\label{EoM_phi}
&\frac{r}{r-r_{0}}\,\partial _t^2
A_{\varphi}-\partial_{r}\left(\frac{r-r_0}{r}\left(\partial_r
A_{\varphi}-\partial_{\varphi}A_{r}\right)\right)
-\frac{\sin\theta}{r^{2}}\partial_{\theta}\left(\frac{1}{\sin\theta}\left(\partial_{\theta}A_{\varphi}-\partial_{\varphi}A_{\theta}\right)\right)=0.
\end{align}
It is the system of equations \eqref{gauge_Sch}--\eqref{EoM_phi}
that will be solved in the next section.

\section{Spectrum of stationary states}
Taking into account the spherical symmetry of equations
\eqref{gauge_Sch}--\eqref{EoM_phi}, we look for stationary
solutions to them characterized by energy $E$, total angular
momentum $j$ and its projection $m$, in the form
\begin{equation} \label{anzats}
\vec A_{jm}\left(E,t,r,\theta ,\varphi\right)
=e^{-iEt}\sum\limits_{\lambda =-1,0,1} {F_j^{(\lambda )} \left(
{E,r} \right)\vec Y_{jm}^{(\lambda )} \left( {\theta ,\varphi }
\right)} ,
\end{equation}
where $\vec A=(A_r,A_\theta,A_\varphi)$ and $\vec Y_{jm
}^{(\lambda )}\left(\theta ,\varphi \right)$ are the spherical
vectors \cite{AB,Rosa:2011my}, which form a complete orthogonal
system of functions in the space of complex vectors on the sphere
$S^2$. For $j\neq 0$, the components of spherical vectors are
written in terms of the ordinary spherical functions $Y_{jm }
\left( {\theta ,\varphi } \right)$ as follows:
\begin{align}\label{spherical_vectors1}
\vec Y_{jm }^{( - 1)} \left( {\theta ,\varphi } \right)&= \left(
{1, 0, 0} \right)Y_{jm } \left( {\theta ,\varphi } \right),
\\\label{spherical_vectors2}
\vec Y_{jm }^{(0)} \left( {\theta
,\varphi } \right)&= \frac{{i}}{{\sqrt {j\left( {j + 1} \right)}
}}\left( {0, \frac{1}{{\sin \theta }}\,\partial _\varphi  ,  -
\sin \theta \, \partial _\theta  } \right)Y_{jm } \left( {\theta
,\varphi } \right), \\\label{spherical_vectors3}
\vec Y_{jm }^{(1)} \left( {\theta ,\varphi } \right)&= \frac{1}{{\sqrt
{j\left( {j + 1}\right)} }}\left( {0, \partial _\theta,\partial
_\varphi} \right)Y_{jm } \left( {\theta ,\varphi } \right).
\end{align}
For $j=0$ there are no spherical vectors with $\lambda=0,1$, and
only $\vec Y_{00}^{(-1)}$ enters the complete system of functions
\cite{AB}. Here the functions
\begin{equation}\label{Ylm}
\begin{gathered}
Y_{jm}(\theta,\varphi)=\sqrt{\frac{2j+1}{4\pi}}\sqrt{\frac{(j-|m|)!}{(j+|m|)!}}\,P_{j}^{|m|}\left(\cos\theta\right)e^{im\varphi}, \\
j=0,1,2, ... ,\qquad m=0,\pm 1, ... ,\pm j,
\end{gathered}
\end{equation}
are spherical harmonics in the convention of \cite{Korn-Korn}.

Substituting expression \eqref{anzats} into equations
\eqref{gauge_Sch}--\eqref{EoM_phi}, carrying out simple
transformations and introducing the notations
\begin{equation}
r^2 F_j^{( - 1)} \left( {E,r} \right) = G_j \left( {E,r} \right),
\qquad F_j^{(0)} \left( {E,r} \right) = F_j \left( {E,r} \right) ,
\end{equation}
we arrive at the equations for $j \ne 0$:
\begin{equation} \label{radial_eq_1}
 \frac{{{d}G_j \left( {E,r} \right) }}{{{d}r}}  - \frac{{\sqrt {j\left( {j + 1} \right)} }}{1 - \frac{r_0}{r}}F_j^{(1)} \left( {E,r} \right)  = 0,
\end{equation}
\begin{equation} \label{radial_eq_2}
 \frac{d}{{{d}r}} \left[ {\left( {1 - \frac{{r_0 }}{r}} \right)\left( {\frac{{{d}F_j^{(1)} \left( {E,r} \right) }}{{{d}r}}  - \frac{{\sqrt {j\left( {j + 1} \right)} }}{{r^2 }}G_j\left( {E,r} \right) } \right)} \right] + \frac{{E^2 }}{1 - \frac{r_0}{r}}F_j^{(1)} \left( {E,r} \right) = 0,
\end{equation}
\begin{equation} \label{radial_eq_3}
 \frac{d}{{{d}r}} \left[ {\left( {1 - \frac{{r_0 }}{r}} \right) \frac{{{d}G_j \left( {E,r} \right) }}{{{d}r}} } \right] - \frac{{j\left( {j + 1} \right)}}{{r^2 }}G_j \left( {E,r} \right) + \frac{{E^2 }}{1 - \frac{r_0}{r}}G_j \left( {E,r} \right) = 0,
\end{equation}
\begin{equation} \label{radial_eq_4}
 \frac{d}{{{d}r}} \left[ {\left( {1 - \frac{{r_0 }}{r}} \right) \frac{{{d}F_j \left( {E,r} \right) }}{{{d}r}} } \right] - \frac{{j\left( {j + 1} \right)}}{{r^2 }}F_j \left( {E,r} \right) + \frac{{E^2 }}{1 - \frac{r_0}{r}}F_j \left( {E,r} \right) = 0.
\end{equation}
We see that we have three coupled equations for the functions
$G_j(E,r)$ and $F_j^{(1)}(E,r)$, whereas the equation for the
function $F_j(E,r)$ decouples. On can also see that
Eqs.~\eqref{radial_eq_3} and \eqref{radial_eq_4} have the same
form. It is easy to check that equation \eqref{radial_eq_2} holds
automatically provided equations \eqref{radial_eq_1},
\eqref{radial_eq_3} are satisfied. Thus, in order to find the
function $F_j(E,r)$ (and $G_j(E,r)$) we have to solve equation
\eqref{radial_eq_4}. Once it is done, the function
$F_j^{(1)}(E,r)$ is found trivially from \eqref{radial_eq_1}.
Without loss of generality, the functions $F_{j}\left(E,r\right)$
and $G_{j}\left(E,r\right)$ are supposed to be real.

The special case is $j=0$, in which we are left only with
\begin{equation}
A_{r}(t,r,\theta,\varphi)=\frac{e^{-iEt}G_{0}(E,r)}{2\sqrt{\pi}\,r^{2}}.
\end{equation}
For such an ansatz  it follows from Eq.~\eqref{EoM_r} that
$E^{2}G_{0}(E,r)=0$, leading to $G_{0}(E,r)\equiv 0$ for $E\neq
0$. For $E=0$ from Eq.~\eqref{gauge_Sch} one gets
$G_{0}(0,r)\equiv\textrm{const}$. However, for such a solution the
field tensor $F_{\mu \nu}$ vanishes, which means that this
solution is a pure gauge and can be discarded. Thus, from here on
we consider only the case $j \ne 0$.

One can show that there are no nonzero solutions to
Eq.~\eqref{radial_eq_4} for $E = 0$. Indeed, let us multiply this
equation by $F_j\left({0,r}\right)$ and integrate it with respect
to $r$ from $r_0$ to $\infty$, performing integration by parts.
The result reads
\begin{equation}
 \left. {\left( {1 - \frac{{r_0 }}{r}} \right)\frac{{{d}F_j \left( {0,r} \right)}}{{{d}r}}F_j \left( {0,r} \right)} \right|_{r_0 }^\infty  - \int\limits_{r_0 }^\infty  {\left[ {\left( {1 - \frac{{r_0
}}{r}} \right)\left( {\frac{{{d}F_j \left( {0,r}
\right)}}{{{d}r}}} \right)^2  + \frac{{j\left( {j + 1}
\right)}}{{r^2 }}F_j^2 \left( {0,r} \right)} \right]{d}r}  = 0.
\end{equation}
From Eq.~\eqref{radial_eq_4} it follows that
$F_j\left({0,r}\right)\sim 1+\frac{j(j+1)}{r_{0}}(r-r_{0})$ for
$r\to r_{0}$ and $F_j\left({0,r}\right)\sim\frac{1}{r^{j}}$ for
$r\to\infty$. Thus, the surface terms vanish, while the integrand
in the second term is nonnegative, which means that $F_j \left(
{0,r} \right) \equiv 0$. This is in accord with the result of
paper \cite{Bekenstein:1971hc}, where  it was shown that a Schwarzschild
black hole cannot have any exterior static scalar or vector
fields.

\begin{figure}[h]
\centering
\includegraphics[width=0.9\textwidth]{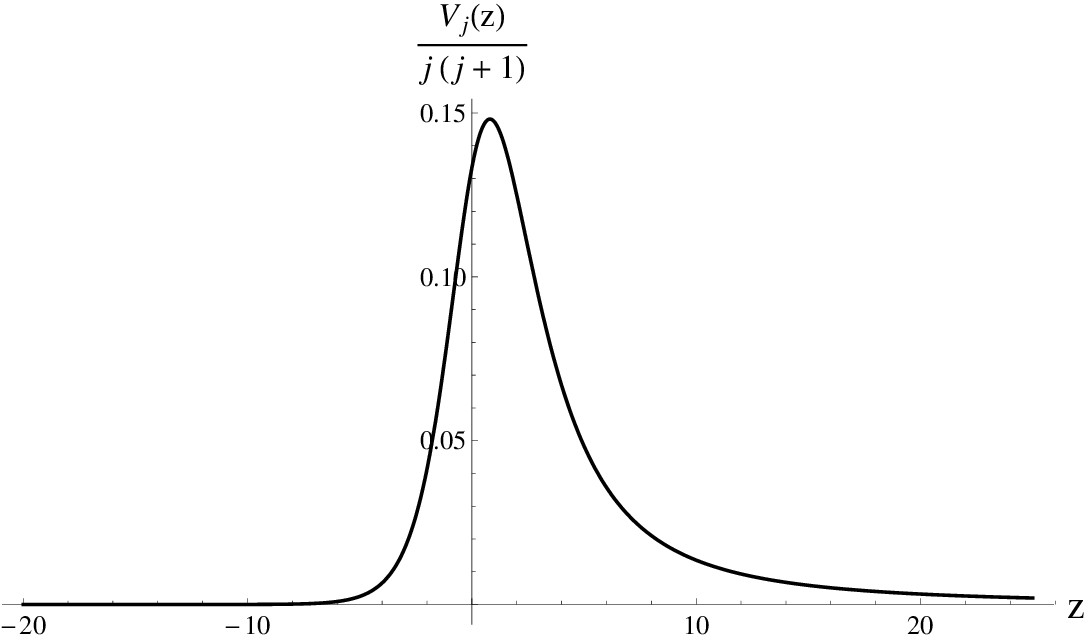}
\caption{Potential $V_{j}(z)$}\label{FigPot}
\end{figure}

In order to study the properties of solutions $F_j(E,r)$, it is
convenient to pass to the tortoise coordinate
\begin{equation}\label{tort}
z=\frac{r}{r_{0}}+\ln\left(\frac{r}{r_{0}}-1\right).
\end{equation}
Then, Eq.~\eqref{radial_eq_4} takes the form of a one-dimensional
Schr\"{o}dinger equation
\begin{equation}\label{EqSchr}
-\frac{d^{2}F_{j}(E,z)}{dz^{2}}+V_{j}(z)\,F_{j}(E,z)=r_{0}^{2}E^{2}\,F_{j}(E,z),
\end{equation}
where
\begin{equation}\label{potSchr}
V_{j}(z)=j(j+1)\frac{\frac{r(z)}{r_{0}}-1}{\left(\frac{r(z)}{r_{0}}\right)^{3}}.
\end{equation}
A remarkable feature of the potential $V_{j}(z)$ is that it
depends on $j$ only through the overall factor $j(j+1)$. Potential
\eqref{potSchr} is presented in Figure~\ref{FigPot}. Equation
\eqref{EqSchr} with potential \eqref{potSchr} coincides with the
equation for the radial parts of physical modes obtained in
\cite{Crispino:2000jx} and with the equation for the radial part
of effective scalar field obtained in \cite{Cognola:1997xp}.

From Eq.~\eqref{EqSchr} it follows that the energy spectrum is
continuous, $0<E<\infty$, and for each $E$ and $j$ there are two
different orthogonal solutions. We will label these solutions by
the index $p=1,2$, i.e., the solutions of equation
\eqref{radial_eq_4} are $F_{jp}(E,r)$.\footnote{Even though the
functions $F_{jp}(E,z)$ are supposed to be real in our approach,
in principle they can be chosen to be complex. For example, in
\cite{Crispino:2000jx} these solutions correspond to ingoing and
outgoing waves.}

Although Eqs.~\eqref{radial_eq_3} and \eqref{radial_eq_4} have the
same form, the functions $F_{jp}(E,r)$ and $G_{jp}(E,r)$ can have
different norms. We suppose that the functions $F_{jp}(E,r)$
satisfy the following normalization condition, which is suggested
by the form of Eq.~\eqref{radial_eq_4}:
\begin{equation} \label{norm}
\int {\frac{{{d}r}}{1 - \frac{r_0}{r}}F_{jp} \left( {E,r}
\right)F_{jp'} \left( {E',r} \right)}  = \delta _{pp'}\,\delta
\left( {E - E'} \right).
\end{equation}
 As for the functions $G_{jp}(E,r)$, we can write
\begin{equation}
G_{jp} \left(E,r\right)=c_{jp}(E)\,F_{jp}\left(E,r\right).
\end{equation}
The coefficients $c_{jp}(E)$ are still unknown and will be defined
later. The completeness relation for the functions $F_{jp}(E,r)$
looks like
\begin{equation} \label{completeness}
\sum\limits_{p=1}^{2}\int {F_{jp} \left( {E,r} \right)F_{jp}
\left( {E,r'} \right){d}E}  = \left( {1 - \frac{{r_0 }}{r'}}
\right)\delta \left( {r - r'} \right).
\end{equation}

Thus, for fixed $E$, $j$, $m$, and $p$ we have two independent
solutions of Eqs.~\eqref{gauge_Sch}--\eqref{EoM_phi}, which look
as follows:
\begin{equation}
e^{-iEt} \vec A_{jmp}^{(a)} \left( {E,\vec r} \right) =
e^{-iEt}c_{jp}(E)\left( {\begin{array}{*{20}c}
   {\frac{1}{{r^2 }}F_{jp} \left( {E,r} \right)Y_{jm} \left( {\theta ,\varphi } \right)}  \\
   \frac{1}{j(j+1)}{\left( {1 - \frac{{r_0 }}{r}} \right)\partial_{r}F_{jp}\left(E,r\right)\partial _\theta  Y_{jm} \left( {\theta ,\varphi } \right)}  \\
    \frac{1}{j(j+1)}{\left( {1 - \frac{{r_0 }}{r}} \right)\partial_{r}F_{jp}\left(E,r\right)\partial _\varphi  Y_{jm} \left( {\theta ,\varphi } \right)}  \\
\end{array}} \right),
\end{equation}
\begin{equation}
e^{-iEt}\vec A_{jmp}^{(b)} \left( {E, \vec r } \right)
=e^{-iEt}\frac{{i}}{{\sqrt {j\left( {j + 1} \right)} }} \, F_{jp}
\left( {E,r} \right)\left( {\begin{array}{*{20}c}
   0  \\
   {\frac{{\rm 1}}{{\sin \theta }}\,\partial _\varphi  Y_{jm} \left( {\theta ,\varphi } \right)}  \\
   { - \sin \theta \,\partial _\theta  Y_{jm} \left( {\theta ,\varphi } \right)}  \\
\end{array}} \right)
\end{equation}
where for convenience, unlike formulas
\eqref{spherical_vectors1}--\eqref{spherical_vectors3}, we write
vectors as columns rather than rows and, for brevity, here and in
some bulky formulas below the arguments $r,\theta,\varphi$ are
replaced by the short notation $\vec r$.

\section{Canonical quantization}
\subsection{Field expansion}
Using the results of the previous section, let us expand the
electromagnetic field in the form
\begin{equation} \label{decomp}
\begin{split}
 \vec A\left( {t, \vec r} \right) = \sum\limits_{p = 1}^2\sum\limits_{j = 1}^\infty\sum\limits_{m =  - j}^j \int\limits_0^\infty  \frac{dE}{\sqrt{2E}} & \left( {e^{ - {i}Et} \vec A_{jmp}^{(a)} \left( {E, \vec r} \right)a_{jmp} \left( E \right) + e^{{i}Et} \vec A_{jmp}^{(a) * } \left( {E, \vec r} \right)a_{jmp}^\dag  \left( E \right)}\right. \\
 & \left. { + \, e^{ - {i}Et} \vec A_{jmp}^{(b)} \left( {E, \vec r} \right)b_{jmp} \left( E \right) + e^{{i}Et} \vec A_{jmp}^{(b)
* } \left( {E, \vec r} \right)b_{jmp}^\dag  \left( E \right)}
\right),
\end{split}
\end{equation}
where the creation and annihilation operators satisfy the
commutation relations
\begin{equation} \label{commutators}
\left[ {a_{jmp} \left( E \right),a_{j'm'p'}^\dag  \left( {E'}
\right)} \right] = \left[ {b_{jmp} \left( E
\right),b_{j'm'p'}^\dag  \left( {E'} \right)} \right] = \delta
_{pp'} \, \delta _{jj'} \, \delta _{mm'} \, \delta \left( {E - E'}
\right),
\end{equation}
all other commutators being equal to zero.

\subsection{Hamiltonian}
First, let us calculate the Hamiltonian of the system, which will
also allow us to find the coefficients $c_{jp}(E)$.
To this end, we use the standard form of the energy-momentum
tensor of the electromagnetic field \cite{BD}
\begin{equation}
T_{\mu \nu}=\frac{1}{4}F_{\rho\sigma}F^{\rho\sigma}g_{\mu
\nu}-F_{\mu \rho}F_{\nu\sigma}g^{\rho\sigma },
\end{equation}
leading to
\begin{equation} \label{Ttt}
T_{tt}  =  - \frac{1}{2}\sum\limits_{k} {\left( {\partial _t A_k }
\right)^2 g^{kk} }  + \frac{1}{2} \sum\limits_{k}\sum\limits_{l}
{\partial _k A_l \left( {\partial _k A_l  - \partial _l A_k }
\right)g^{kk} g^{ll} g_{tt} },
\end{equation}
where $k,l = r,\theta,\varphi$. By definition, the Hamiltonian is
given by
\begin{equation}
H = \int {T_{tt} \,g^{tt} \sqrt { - g} \,{d}^3 x} .
\end{equation}
Substituting \eqref{Ttt} into this formula, integrating by parts
and making use of  equations of motion \eqref{EoM}, we obtain
\begin{equation} \label{H}
H =\frac{1}{2}\int{\sum\limits_k {\left[A_k \partial_t^2
A_k-\left({\partial_t A_k }\right)^2\right]g^{kk}}g^{tt}\sqrt{-g}
\,{d}^3 x} .
\end{equation}

Substituting \eqref{decomp} into expression \eqref{H} and
performing straightforward but rather bulky calculations (they are
presented in Appendix~C), one gets the explicit form of
the coefficients $c_{jp}(E)$, which are given by
\begin{equation}\label{coeff-c}
c_{jp}(E)=\frac{\sqrt{j(j+1)}}{E},
\end{equation}
and the Hamiltonian
\begin{equation}\label{H-final}
H = \sum\limits_{p = 1}^2 {\sum\limits_{j = 1}^\infty
{\sum\limits_{m =  - j}^j {\int\limits_0^\infty  {{d}E\,E\left(
{a_{jmp}^\dag  \left( E \right)a_{jmp} \left( E \right) +
b_{jmp}^\dag  \left( E \right)b_{jmp} \left( E \right)} \right)} }
} } .
\end{equation}
In this Hamiltonian, we passed to the normal ordering and dropped
infinite $c$-number terms. This Hamiltonian has the form that is
expected for the Hamiltonian of electromagnetic field in Minkowski
spacetime in spherical coordinates, except for the doubling of
number of quantum states. An analogous doubling was found in the
cases of scalar \cite{Egorov:2022hgg,Smolyakov:2023pml} and spinor
\cite{Egorov:2023ruz} fields.

\subsection{Canonical commutation relations}
Having found coefficients \eqref{coeff-c}, we are ready to
calculate the canonical commutation relations. The canonical
coordinates in the theory are the set of fields $A_k \left(
{t,\vec r} \right)$, and, by definition, the canonically
conjugate momenta are
\begin{equation}
\pi ^l \left( {t,\vec r} \right) =  - \sqrt { - g\left( {\vec r}
\right)} \, g^{tt} \left( {\vec r} \right)g^{ll} \left( {\vec r}
\right)\dot A_l \left( {t,\vec r} \right)
\end{equation}
(of course, summation with respect to $l$ is not implied here),
$k,l = r,\theta ,\varphi$. We expect that the following canonical
commutation relations should be satisfied:
\begin{align} \label{commutator_xp}
&\left[ {A_k \left( {t,\vec r} \right),\pi ^l \left( {t,\vec r^{\, \prime}} \right)} \right] = {i}D_k^l \left( {\vec r, \vec r^{\, \prime}} \right),
\\\label{commutator_xx}
&\left[ {A_k \left( {t,\vec r} \right),A_l \left( {t,\vec r^{\, \prime}} \right)} \right] = 0,
\\\label{commutator_pp}
&\left[ {\pi ^k \left( {t,\vec r} \right),\pi ^l \left( {t,\vec r^{\, \prime}} \right)} \right] = 0,
\end{align}
where $D_k^l \left( {\vec r, \vec r^{\, \prime}} \right)$ is a
generalized function such that commutation relations
\eqref{commutator_xp} are consistent with gauge condition
\eqref{gauge_Sch}. It is difficult to predict the form of $D_k^l
\left( {\vec r, \vec r^{\, \prime}} \right)$ in the Schwarzschild
spacetime, so we will derive $D_k^l \left( {\vec r, \vec
r^{\, \prime}} \right)$ directly by calculating the lhs of
\eqref{commutator_xp} for field expansion \eqref{decomp}.

First, let us consider commutation relations
\eqref{commutator_xx}. We start with the simplest example $k=l=r$.
After straightforward calculations, making use of commutators
\eqref{commutators}, one gets
\begin{align}\nonumber
 \left[ {A_r \left( {t,\vec r} \right),A_r \left( {t,\vec r^{\, \prime}} \right)} \right]& = \sum\limits_{p = 1}^2 {\sum\limits_{j = 1}^\infty  {\sum\limits_{m =  - j}^j {\int\limits_0^\infty  {\frac{{{d}E}}{{2E^3 }}\frac{{j\left( {j + 1} \right)}}{{r^2 r'^2 }}\,F_{jp} \left( {E,r} \right)F_{jp} \left( {E,r'} \right)}}}}  \\\label{Ar_Ar}
 &\times \Bigl(Y_{jm}\left(\theta,\varphi\right)Y_{jm}^*\left(\theta',\varphi'\right)-Y_{jm}^*\left(\theta ,\varphi\right)Y_{jm}\left(\theta',\varphi'\right) \Bigl).
\end{align}
Recall that there exists the addition theorem for spherical
harmonics \cite{Korn-Korn}:
\begin{equation} \label{add}
\sum\limits_{m =  - j}^j {Y_{jm} \left( {\theta ,\varphi }
\right)Y_{jm}^ *  \left( {\theta ',\varphi '} \right)}  =
\frac{{2j + 1}}{{4\pi }}P_j \left( {\cos \omega } \right),
\end{equation}
where $\cos \omega  = \cos \theta \cos \theta ' + \sin \theta \sin
\theta '\cos \left( {\varphi  - \varphi '} \right)$ (i.e.,
$\omega$ is the angle between rays $\left( {\theta ,\varphi }
\right)$ and $\left( {\theta ',\varphi '} \right)$ in the
spherical coordinate system). By virtue of \eqref{add}, the
construction in the brackets in the rhs of \eqref{Ar_Ar}
immediately vanishes:
\begin{align}\nonumber
\sum\limits_{m =  - j}^j \Bigl(Y_{jm}\left(\theta,\varphi\right)Y_{jm}^*\left(\theta',\varphi'\right)-Y_{jm}^*\left(\theta ,\varphi\right)Y_{jm}\left(\theta',\varphi'\right) \Bigl)&\\
=\frac{{2j + 1}}{{4\pi }}P_j \left( {\cos \omega }
\right)-\frac{{2j + 1}}{{4\pi }}P_j \left( {\cos \omega }
\right)&= 0.
\end{align}
It is easy to check that an analogous cancellation occurs in all
the other commutation relations \eqref{commutator_xx}, and also in
the commutation relations $\big[ {\dot A_k \left( {t,\vec r}
\right), \dot A_l \left( {t,\vec r^{\, \prime}} \right)} \big]$,
which leads to \eqref{commutator_pp}.

Now let us calculate commutation relations \eqref{commutator_xp}.
Again we start with the simplest case $k=l=r$. Making use of
commutators \eqref{commutators}, as a result of straightforward
calculations we obtain
\begin{align}\nonumber
 \left[ {A_r \left( {t,\vec r} \right),\dot A_r \left( {t,\vec r^{\, \prime}} \right)} \right] &= {i}\sum\limits_{p = 1}^2 {\sum\limits_{j = 1}^\infty  {\sum\limits_{m =  - j}^j {\int\limits_0^\infty  {\frac{{{d}E}}{{2E^2 }}\frac{{j\left( {j + 1} \right)}}{{r^2 r'^2 }}\,F_{jp} \left( {E,r} \right)F_{jp} \left( {E,r'} \right)} } } } \\\label{Ar_dotAr_1}
 & \times \Bigl(Y_{jm} \left( {\theta ,\varphi } \right)Y_{jm}^*  \left( {\theta ',\varphi '} \right)+Y_{jm}^*  \left( {\theta ,\varphi } \right)Y_{jm} \left( {\theta ',\varphi '} \right) \Bigr).
\end{align}
The spherical functions satisfy the equation
\begin{equation}\label{Y_eq}
\Delta_{\theta,\varphi}Y_{jm} \left( {\theta ,\varphi } \right) +
j\left( {j + 1} \right)Y_{jm} \left( {\theta ,\varphi } \right) =
0,
\end{equation}
where $\Delta _{\theta ,\varphi}$ is the angular part of the
Laplace operator. Taking into account that the functions $F_{jp}
\left( {E,r} \right)$ are determined by eigenvalue problem
\eqref{radial_eq_4}, we can introduce the operator $Q$, which acts
on an arbitrary function $g(r,\theta,\varphi)$ as
\begin{equation}
Q g(r,\theta,\varphi)\equiv \left( {1 - \frac{{r_0 }}{r}}
\right)\partial_{r}\left[ {\left( {1 - \frac{{r_0 }}{r}}
\right)\partial_{r}g(r,\theta,\varphi)} \right] + \left( {1 -
\frac{{r_0 }}{r}} \right)\frac{\Delta_{\theta
,\varphi}}{r^2}\,g(r,\theta,\varphi).
\end{equation}
Then the product $F_{jp} \left( {E,r} \right)Y_{jm} \left( {\theta
,\varphi } \right)$ satisfies the equivalent eigenvalue problem
\begin{equation}
Q\left[ {F_{jp} \left( {E,r} \right)Y_{jm} \left( {\theta ,\varphi
} \right)} \right] + E^2 F_{jp} \left( {E,r} \right)Y_{jm} \left(
{\theta ,\varphi } \right) = 0 .
\end{equation}
Thus, one can write
\begin{equation}
\frac{1}{{E^2 }}F_{jp} \left( {E,r} \right)Y_{jm} \left( {\theta
,\varphi } \right) =  - \frac{1}{Q}\left[ {F_{jp} \left( {E,r}
\right)Y_{jm} \left( {\theta ,\varphi } \right)} \right],
\end{equation}
and commutator \eqref{Ar_dotAr_1} can be brought to the form
\begin{align}\nonumber
 \left[ {A_r \left( {t,\vec r} \right),\dot A_r \left( {t,\vec r^{\, \prime}} \right)} \right]& = \frac{{i}}{2}\frac{1}{{r^2 r'^2 }}\frac{{\Delta_{\theta ,\varphi } }}{Q}\sum\limits_{p = 1}^2 {\sum\limits_{j = 1}^\infty  {\sum\limits_{m =  - j}^j {\int\limits_0^\infty  {{d}E \,F_{jp} \left( {E,r} \right)F_{jp} \left( {E,r'} \right)} } } } \ \ \\\label{Ar_dotAr_2}
 &\times \Bigl(Y_{jm} \left( {\theta ,\varphi } \right)Y_{jm}^*  \left( {\theta ',\varphi '} \right)+Y_{jm}^*  \left( {\theta ,\varphi } \right)Y_{jm} \left( {\theta ',\varphi '} \right) \Bigr).
\end{align}
Making use of addition theorem \eqref{add} and completeness
relation \eqref{completeness}, we rewrite expression
\eqref{Ar_dotAr_2} in the form
\begin{equation}
\left[ {A_r \left( {t,\vec r} \right),\dot A_r \left( {t,\vec
r^{\, \prime}} \right)} \right] = \frac{{i}}{4\pi}\frac{1}{{r^2 r'^2
}}\frac{{\Delta _{\theta ,\varphi } }}{Q}\left[ {\left( {1 -
\frac{{r_0 }}{r'}} \right)\delta \left( {r - r'}
\right)\sum\limits_{j = 1}^\infty  {\left( {2j + 1} \right)P_j
\left( {\cos \omega } \right)} } \right] .
\end{equation}
Given that \cite{LL-QM}
\begin{equation}
\frac{1}{4}\sum\limits_{j = 0}^\infty  {\left( {2j + 1} \right)P_j
\left( {\cos \omega } \right)}  = \delta \left( {1 - \cos \omega }
\right),
\end{equation}
which means that
\begin{equation}
\sum\limits_{j = 1}^\infty  {\left( {2j + 1} \right)P_j \left(
{\cos \omega } \right)}  =
4\left(\delta\left(1-\cos\omega\right)-\frac{1}{4}\right),
\end{equation}
we arrive at
\begin{equation}
\left[ {A_r \left( {t,\vec r} \right),\dot A_r \left( {t,\vec
r^{\, \prime}} \right)} \right] = \frac{i}{\pi }\frac{1}{{r^2 r'^2
}}\left( {1 - \frac{{r_0 }}{{r'}}} \right)\frac{{\Delta _{\theta
,\varphi } }}{Q}\left[ {\delta \left( {r - r'}
\right)\left(\delta\left(1-\cos\omega\right)-\frac{1}{4}\right)}
\right]
\end{equation}
and, finally, at
\begin{equation}\label{crel-1}
\left[ {A_r \left( {t,\vec r} \right),\pi ^r \left( {t,\vec
r^{\, \prime}} \right)} \right] =\frac{{i\sin \theta '}}{{\pi r^2
}}\left( {1 - \frac{{r_0 }}{{r'}}} \right)\frac{{\Delta _{\theta
,\varphi } }}{Q}\left[ {\delta \left( {r - r'}
\right)\left(\delta\left(1-\cos\omega\right)-\frac{1}{4}\right)}
\right].
\end{equation}

In exactly the same manner but with slightly more complicated
calculations one can obtain the remaining canonical commutation
relations \eqref{commutator_xp}:
\begin{equation}
\left[ {A_r \left( {t,\vec r} \right),\pi ^\theta  \left( {t,\vec
r^{\, \prime}} \right)} \right] =  -\frac{{i\sin \theta '}}{{\pi r^2
}}\frac{{\partial _{r'} \partial _{\theta '} }}{Q}\left[ {\left(
{1 - \frac{{r_0 }}{{r'}}} \right)\delta \left( {r - r'}
\right)\left(\delta\left(1-\cos\omega\right)-\frac{1}{4}\right)}
\right],
\end{equation}

\begin{equation}
\left[ {A_r \left( {t,\vec r} \right),\pi ^\varphi  \left(
{t,\vec r^{\, \prime}} \right)} \right] =  - \frac{i}{{\pi r^2 \sin
\theta '}}\frac{{\partial _{r'} \partial _{\varphi '} }}{Q}\left[
{\left( {1 - \frac{{r_0 }}{{r'}}} \right)\delta \left( {r - r'}
\right)\left(\delta\left(1-\cos\omega\right)-\frac{1}{4}\right)}
\right],
\end{equation}

\begin{equation}
\left[ {A_\theta  \left( {t,\vec r} \right),\pi ^r \left( {t,\vec r^{\, \prime}} \right)} \right] =  - \frac{i\sin\theta'}{\pi }\left( {1 - \frac{{r_0 }}{r}} \right)\left( {1 - \frac{{r_0 }}{{r'}}} \right)
 \frac{\partial _r \partial _\theta}{Q}\left[ {\delta
\left( {r - r'}
\right)\left(\delta\left(1-\cos\omega\right)-\frac{1}{4}\right)}
\right] ,
\end{equation}

\begin{equation}
\begin{split}
 & \left[ {A_\theta  \left( {t,\vec r} \right),\pi ^\theta  \left( {t,\vec r^{\, \prime}} \right)} \right] =  - \frac{i}{\pi }\left\{ {\frac{1}{{\sin \theta }}\,\delta \left( {r - r'} \right)\frac{{\partial _\varphi  \partial _{\varphi '} }}{{\Delta _{\theta ,\varphi } }}\,\left(\delta\left(1-\cos\omega\right)-\frac{1}{4}\right)}\,\right. \\
 &  \left. { - \sin\theta'\left( {1 - \frac{{r_0 }}{r}} \right)\frac{\partial _r \partial _\theta{\partial _{r'} \partial _{\theta '} }}{{\Delta _{\theta ,\varphi } Q}}\left[ {\left( {1 - \frac{{r_0 }}{{r'}}} \right)\delta \left( {r - r'} \right)\left(\delta\left(1-\cos\omega\right)-\frac{1}{4}\right)} \right]} \right\},
\end{split}
\end{equation}

\begin{equation}
\begin{split}
 & \left[ {A_\theta  \left( {t,\vec r} \right),\pi ^\varphi  \left( {t,\vec r^{\, \prime}} \right)} \right] = \frac{i}{\pi }\left\{ {\frac{1}{{\sin \theta }}\,\delta \left( {r - r'} \right)\frac{{\partial _\varphi  \partial _{\theta '} }}{{\Delta _{\theta ,\varphi } }}\,\left(\delta\left(1-\cos\omega\right)-\frac{1}{4}\right)} \,\right. \\
 & \left. { + \frac{1}{{\sin\theta'}}\left( {1 - \frac{{r_0 }}{r}} \right)\frac{\partial _r \partial _\theta{\partial _{r'} \partial _{\varphi '} }}{{\Delta _{\theta ,\varphi } Q}}\left[ {\left( {1 - \frac{{r_0 }}{{r'}}} \right)\delta \left( {r - r'} \right)\left(\delta\left(1-\cos\omega\right)-\frac{1}{4}\right)} \right]} \right\},
\end{split}
\end{equation}

\begin{equation}
\left[ {A_\varphi  \left( {t,\vec r} \right),\pi ^r \left( {t,\vec r^{\, \prime}} \right)} \right] =  - \frac{i\sin\theta'}{\pi }\left( {1 - \frac{{r_0 }}{r}} \right)\left( {1 - \frac{{r_0 }}{{r'}}} \right)
 \frac{\partial _r{\partial _\varphi  }}{Q}\left[ {\delta
\left( {r - r'}
\right)\left(\delta\left(1-\cos\omega\right)-\frac{1}{4}\right)}
\right],
\end{equation}

\begin{equation}
\begin{split}
 & \left[ {A_\varphi  \left( {t,\vec r} \right),\pi ^\theta  \left( {t,\vec r^{\, \prime}} \right)} \right] = \frac{i}{\pi }\left\{ {\sin \theta \, \delta \left( {r - r'} \right) \frac{\partial _\theta{\partial _{\varphi '} }}{{\Delta _{\theta ,\varphi } }} \, \left(\delta\left(1-\cos\omega\right)-\frac{1}{4}\right)} \,\right. \\
 & \left. { + \sin\theta'\left( {1 - \frac{{r_0 }}{r}} \right)\frac{\partial _r{\partial _{r'} \partial _\varphi  \partial _{\theta '} }}{{\Delta _{\theta ,\varphi } Q}}\left[ {\left( {1 - \frac{{r_0 }}{{r'}}} \right)\delta \left( {r - r'} \right)\left(\delta\left(1-\cos\omega\right)-\frac{1}{4}\right)} \right]} \right\},
\end{split}
\end{equation}

\begin{equation}
\begin{split}
 & \left[ {A_\varphi  \left( {t,\vec r} \right),\pi ^\varphi  \left( {t,\vec r^{\, \prime}} \right)} \right] =  - \frac{i}{\pi }\left\{ {\sin \theta \,\delta \left( {r - r'} \right)\frac{\partial _\theta{\partial _{\theta '} }}{{\Delta _{\theta ,\varphi } }} \, \left(\delta\left(1-\cos\omega\right)-\frac{1}{4}\right)} \,\right. \\\label{crel-9}
 & \left. { - \frac{1}{{\sin\theta'}}\left( {1 - \frac{{r_0 }}{r}} \right)\frac{\partial _r{\partial _{r'} \partial _\varphi  \partial _{\varphi '} }}{{\Delta _{\theta ,\varphi } Q}}\left[ {\left( {1 - \frac{{r_0 }}{{r'}}} \right)\delta \left( {r - r'} \right)\left(\delta\left(1-\cos\omega\right)-\frac{1}{4}\right)} \right]} \right\}.
\end{split}
\end{equation}
One can see that relations \eqref{crel-1}--\eqref{crel-9} are
quite complicated and do not resemble rather simple commutation
relations in the case of electromagnetic field in the Coulomb
gauge in Minkowski spacetime. However, such a complexity is
inherent not only to the case of the Schwarzschild spacetime.
Indeed, by setting $r_{0}=0$ in \eqref{crel-1}--\eqref{crel-9}, we
get the commutation relations for electromagnetic field in the
Coulomb gauge in spherical coordinates in Minkowski spacetime. One
can see that even in this case the commutation relations have a
rather complicated form.

\section{Conclusion}
In the present paper, we have carried out the procedure of
canonical  quantization of electromagnetic field above the horizon
of an ideal Schwarzschild black hole. It is shown that in such a
setup a consistent quantum theory of the electromagnetic field can
be constructed in complete analogy with the cases of scalar
\cite{Egorov:2022hgg,Smolyakov:2023pml} and spinor
\cite{Egorov:2023ruz} fields. Note that the interior of the black
hole (i.e., the area below the horizon) does not affect the
resulting quantum field theory, exactly as in
\cite{Egorov:2022hgg,Smolyakov:2023pml,Egorov:2023ruz}. The
Hamiltonian of the theory has the standard form without any
pathologies. Note that the results presented in this paper can
also be applied (by setting $r_{0}=0$ and removing the doubling of
the radial solutions) to the case of electromagnetic field in
Minkowski spacetime that is quantized in spherical coordinate
system (we have failed to find a discussion of such a case in the
literature).

Contrary to the cases of massive scalar and spinor fields,  here
exists only one branch of quantum states. The branch of states
with energies smaller than the field mass, which corresponds to
particles trapped in the vicinity of the horizon, is absent due to
the mass of electromagnetic field being equal to zero. Thus, here
we have a continuous spectrum of states with energies starting
from zero corresponding to the infinite motion of particles. The
states in the spectrum are doubly degenerate. As was noted above,
there is no such a degeneracy of states in the case of
electromagnetic field in spherical coordinates in Minkowski
spacetime. As was explicitly demonstrated in
\cite{Egorov:2022hgg}, this degeneracy is due to the topological
structure $\mathbb{R}^2 \times S^2$ of the Schwarzschild
spacetime, which differs from the topological structure
$\mathbb{R}^4$ of Minkowski spacetime. Thus, the appearance of
analogous degeneracies is expected in other curved spacetimes like
the one of wormholes (see discussion in \cite{Egorov:2022hgg}) or
black holes of other types.

As in the case of spinor field \cite{Egorov:2023ruz}, the above
procedure of canonical quantization of electromagnetic field is
based on the use of solutions to the equations of motion in the
Schwarzschild coordinates. However, one can expect that, as in the
case of scalar field \cite{Egorov:2022hgg,Smolyakov:2023pml},
there is a way to pass to some linear combinations of the
scatteringlike states \cite{Egorov:2022hgg,Smolyakov:2023pml},
which are more convenient for describing the theory at large
distances from the Schwarzschild black hole. Indeed, although the
case of electromagnetic field is technically much more complicated
than the scalar field case and so the corresponding steps are not
clear at the moment, the structure of the spectrum of
electromagnetic field (namely, the degeneracy of the radial
solutions) is similar to the structure of the spectrum of the
massless scalar field \cite{Egorov:2022hgg}.

The construction of quantum theories of the spinor and
electromagnetic fields in the Schwarzschild spacetime allows us to
move on to quantum electrodynamics in this spacetime, which makes
it possible to consider a quantum description of the fall of
charged particles towards the black hole with the emission of
photons. Such a description taking into account the doubling of
quantum states could affect the results concerning the rate of
accretion of matter by the black hole or predict other observable
effects. This problem is the subject of further research.

\section*{Acknowledgements}
This study was conducted within the scientific program of the
National Center for Physics and Mathematics, section \textnumero 5
``Particle Physics and Cosmology.'' Stage 2023-2025.

\section*{Appendix~A: Gauge condition}
\begin{figure}[H]
\centering
\begin{minipage}{.49\textwidth}
\centering
\includegraphics[width=0.95\linewidth]{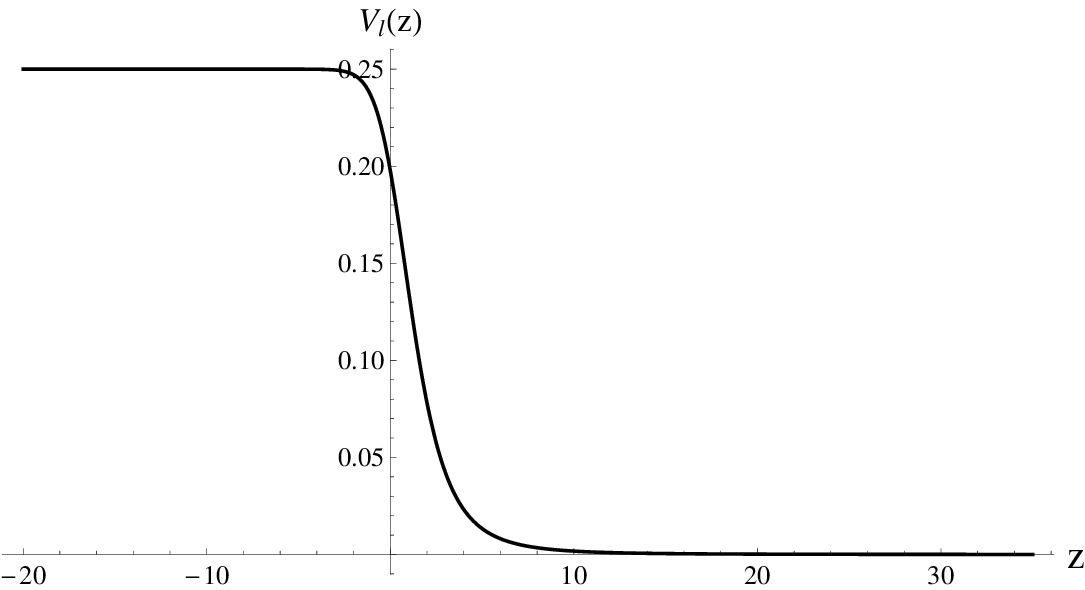}
\end{minipage}
\begin{minipage}{.49\textwidth}
\centering
\includegraphics[width=0.95\linewidth]{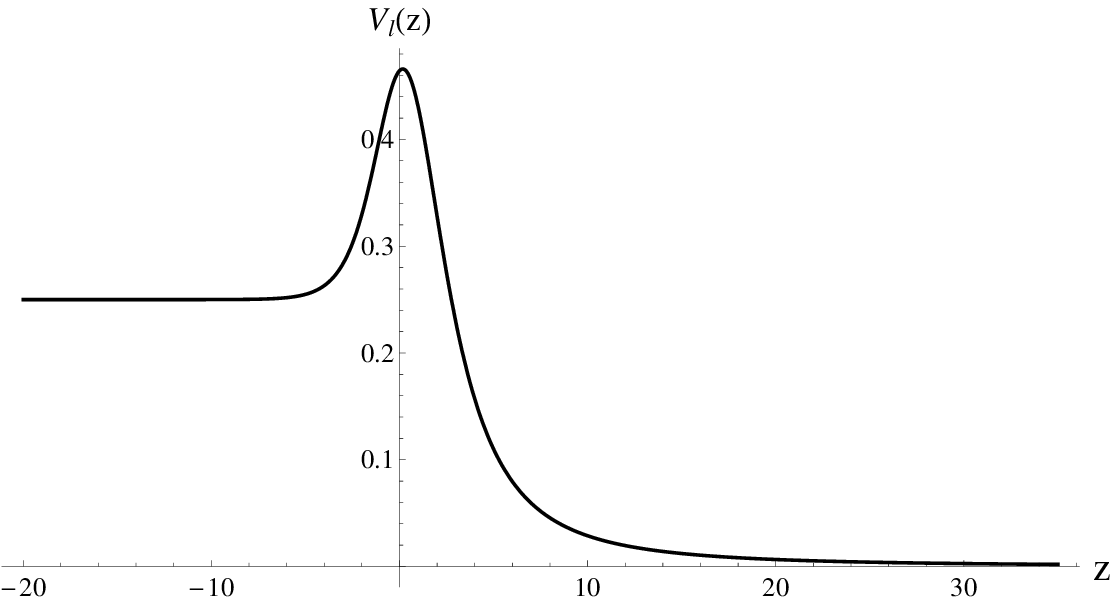}
\end{minipage}
\caption{$V_{l}(z)$ for $l=0$ (left plot) and $l=1$ (right
plot)}\label{pot-gauge-fig}
\end{figure}

\setcounter{equation}{0}
\renewcommand{\theequation}{A\arabic{equation}}
Suppose that
\begin{equation}
\nabla^{\mu}{A'}_{\mu}+\frac{{r_0 }}{{r^2 }}A'_{r}=0.
\end{equation}
Then, for the gauge function $\alpha(r,\theta,\varphi)$ defined by
\begin{equation}
A'_{k}(t,r,\theta,\varphi)=A_{k}(t,r,\theta,\varphi)-\partial_{k}\alpha(r,\theta,\varphi)
\end{equation}
we get the equation
\begin{equation}
\left(1-\frac{r_{0}}{r}\right)\left(\frac{\partial^{2}\alpha}{\partial
r^{2}}+\frac{2}{r}\frac{\partial\alpha}{\partial
r}\right)+\frac{1}{r^{2}}\Delta_{\theta
,\varphi}\alpha=f(r,\theta,\varphi).
\end{equation}
Let us expand the functions $\alpha(r,\theta,\varphi)$ and
$f(r,\theta,\varphi)$ in spherical harmonics as
\begin{align}
&\alpha(r,\theta,\varphi)=\sum\limits_{l=0}^{\infty}\sum\limits_{m=-l}^{l}Y_{lm}\left({\theta ,\varphi}\right)\alpha_{lm}(r),\\
&f(r,\theta,\varphi)=\sum\limits_{l=0}^{\infty}\sum\limits_{m=-l}^{l}Y_{lm}\left({\theta
,\varphi}\right)f_{lm}(r).
\end{align}
Then, we get the equations
\begin{equation}
\left(1-\frac{r_{0}}{r}\right)\left(\frac{d^{2}\alpha_{lm}(r)}{dr^{2}}+\frac{2}{r}\frac{d\alpha_{lm}(r)}{dr}\right)-\frac{l(l+1)}{r^{2}}\alpha_{lm}(r)
=f_{lm}(r)
\end{equation}
for the radial functions $\alpha_{lm}(r)$. Passing to the tortoise
coordinate \eqref{tort}, we get the equation
\begin{equation}
\frac{d^{2}\alpha_{lm}(z)}{dz^{2}}+\frac{r_{0}(2r(z)-3r_{0})}{r^{2}(z)}\frac{d\alpha_{lm}(z)}{dz}-\frac{l(l+1)(r(z)-r_{0})r_{0}^{2}}{r^{3}(z)}\,\alpha_{lm}(z)
=r_{0}^{2}\frac{r(z)-r_{0}}{r(z)}\,f_{lm}(z).
\end{equation}
Using the substitution
\begin{equation}
\alpha_{lm}(z)=\frac{\sqrt{\frac{r(z)}{r_{0}}-1}}{\left(\frac{r(z)}{r_{0}}\right)^{\frac{3}{2}}}\,\beta_{lm}(z),
\end{equation}
we finally get the equation
\begin{equation}\label{eqSchr-gauge}
-\frac{d^{2}\beta_{lm}(z)}{dz^{2}}+V_{l}(z)\beta_{lm}(z)
=-r_{0}^{2}\sqrt{\frac{r(z)}{r_{0}}\left(\frac{r(z)}{r_{0}}-1\right)}\,f_{lm}(z)
\end{equation}
with
\begin{equation}\label{pot-gauge}
V_{l}(z)=\frac{4\frac{r(z)}{r_{0}}-3}{4\left(\frac{r(z)}{r_{0}}\right)^{4}}+l(l+1)\frac{\frac{r(z)}{r_{0}}-1}{\left(\frac{r(z)}{r_{0}}\right)^{3}}.
\end{equation}
The potential $V_{l}(z)$ is such that $V_{l}(z)\to\frac{1}{4}$ for
$z\to-\infty$ and $V_{l}(z)\to 0$ for $z\to\infty$. The examples
of potential \eqref{pot-gauge} can be found in
Figure~\ref{pot-gauge-fig}.

In order to find a solution of \eqref{eqSchr-gauge}, one should
consider the auxiliary Schr\"{o}dinger-like equation
\begin{equation}
-\frac{d^{2}\tilde\beta_{l}(\epsilon,z)}{dz^{2}}+V_{l}(z)\tilde\beta_{l}(\epsilon,z)
=\epsilon\tilde\beta_{l}(\epsilon,z).
\end{equation}
Without loss of generality, the functions
$\tilde\beta_{l}(\epsilon,z)$ are supposed to be real. It is clear
that for $\epsilon<\frac{1}{4}$ there exists only one solution
$\tilde\beta_{l}(\epsilon,z)$ for a given $\epsilon$, whereas for
$\epsilon>\frac{1}{4}$ there exist two linearly independent
solutions which will be denoted as $\tilde\beta_{lp}(\epsilon,z)$,
$p=1,2$. If properly normalized, these solutions form a complete
set of orthogonal eigenfunctions, the corresponding completeness
relation having the form
\begin{equation}
\int\limits_{0}^{\frac{1}{4}}\tilde\beta_{l}(\epsilon,z)\tilde\beta_{l}(\epsilon,z')d\epsilon+\sum\limits_{p=1}^{2}
\int\limits_{\frac{1}{4}}^{\infty}\tilde\beta_{lp}(\epsilon,z)\tilde\beta_{lp}(\epsilon,z')d\epsilon=\delta(z-z').
\end{equation}
Thus, the Green's function of the problem \eqref{eqSchr-gauge} is
\begin{equation}
G(z,z')=\int\limits_{0}^{\frac{1}{4}}\frac{\tilde\beta_{l}(\epsilon,z)\tilde\beta_{l}(\epsilon,z')}{\epsilon}\,d\epsilon+\sum\limits_{p=1}^{2}
\int\limits_{\frac{1}{4}}^{\infty}\frac{\tilde\beta_{lp}(\epsilon,z)\tilde\beta_{lp}(\epsilon,z')}{\epsilon}\,d\epsilon,
\end{equation}
whereas the solutions for $\beta_{lm}(z)$ have the form
\begin{equation}
\beta_{lm}(z)=-r_{0}^{2}\int\limits_{-\infty}^{\infty}G(z,z')\sqrt{\frac{r(z')}{r_{0}}\left(\frac{r(z')}{r_{0}}-1\right)}\,f_{lm}(z')dz'.
\end{equation}
Given the solutions $\beta_{lm}(z)$, the solution for the gauge
function $\alpha(r,\theta,\varphi)$ can be easily obtained.

\section*{Appendix~B: Gauge condition in the isotropic coordinates}
\setcounter{equation}{0}
\renewcommand{\theequation}{B\arabic{equation}}
We start with gauge condition \eqref{gauge2}. Let us consider the
term $\frac{r_0}{r^2}A_r$, which can be rewritten as
\begin{align}\nonumber
\frac{r_0}{r^2}A_r
&=\frac{r_{0}}{R^{2}\left(1+\frac{r_{0}}{4R}\right)^4}\,\frac{\partial
R}{\partial r}\,A_{R}
=\frac{r_{0}}{R^{2}\left(1+\frac{r_{0}}{4R}\right)^4}\,\frac{1}{1-\left(\frac{r_{0}}{4R}\right)^{2}}\,A_{R}
\\\label{gaugeisotr2}&=\frac{r_{0}}{R^{2}\left(1+\frac{r_{0}}{4R}\right)^4}\,\frac{1}{1-\left(\frac{r_{0}}{4R}\right)^{2}}
\sum\limits_{i=1}^{3}\frac{\partial X^{i}}{\partial R}A_{i},
\end{align}
where formula \eqref{Rrtransform} was used to calculate
$\frac{\partial R}{\partial r}$. From definition
\eqref{RtransformXYZ} it follows that $\frac{\partial
X^{i}}{\partial R}=\frac{X^{i}}{R}$, resulting in
\begin{equation}\label{gaugeisotr2b}
\frac{r_0}{r^2}A_r=\frac{r_{0}}{R^{3}\left(1+\frac{r_{0}}{4R}\right)^4}\,\frac{1}{1-\left(\frac{r_{0}}{4R}\right)^{2}}\,
(\vec r\vec A).
\end{equation}

In the isotropic coordinates, the term $\nabla^{\mu}A_{\mu}$ in
formula \eqref{gauge2} (where $\mu$ now labels the coordinates
$t$, $X^{1}$, $X^{2}$, $X^{3}$) can be rewritten as \cite{LL-FT}
\begin{equation}\label{nablasigma}
\nabla^{\mu}A_{\mu} =\frac{1}{\sqrt{-g}}\,\partial_{\mu}\left(\sqrt{-g}A^{\mu}\right)=\frac{1}{\sqrt{-g}}\,\partial_{j}\left(\sqrt{-g}g^{ij}A_{i}\right)
 =g^{ij}\partial_{j}A_{i}+\frac{1}{\sqrt{-g}}\,\partial_{j}\left(\sqrt{-g}g^{ij}\right)A_{i},
\end{equation}
where $i,j=1,2,3$. One can check that in the isotropic coordinates
(see, for example, \cite{LL-FT})
\begin{equation}
\sqrt{-g}=\left(1-\left(\frac{r_{0}}{4R}\right)^{2}\right)\left(1+\frac{r_{0}}{4R}\right)^4,
\qquad\sqrt{-g}g^{ij}=-\left(1-\left(\frac{r_{0}}{4R}\right)^{2}\right)\delta_{ij}.
\end{equation}
With the latter relations, formula \eqref{nablasigma} can be
rewritten as
\begin{align}\nonumber
&g^{ij}\partial_{j}A_{i}+\frac{1}{\sqrt{-g}}\,\partial_{j}\left(\sqrt{-g}g^{ij}\right)A_{i}\\\nonumber
&
=-\frac{1}{\left(1+\frac{r_{0}}{4R}\right)^4}\sum\limits_{i=1}^{3}\partial_{i}A_{i}
-\frac{r_{0}^{2}}{8R^{4}\left(1-\left(\frac{r_{0}}{4R}\right)^{2}\right)\left(1+\frac{r_{0}}{4R}\right)^4}
\sum\limits_{i=1}^{3}X^{i}A_{i}\\\label{gaugeisotr1}
&=-\frac{1}{\left(1+\frac{r_{0}}{4R}\right)^4}\,\textrm{div}\vec
A-\frac{r_{0}^{2}}{8R^{4}\left(1-\left(\frac{r_{0}}{4R}\right)^{2}\right)\left(1+\frac{r_{0}}{4R}\right)^4}(\vec
R\vec A),
\end{align}
where the relation
\begin{equation}
\partial_{i}R=\partial_{i}\sqrt{\sum\limits_{i=1}^{3}(X^{i})^{2}}=\frac{X^{i}}{R}
\end{equation}
was used. Adding \eqref{gaugeisotr2b} to \eqref{gaugeisotr1} and
multiplying the result by $-\left(1+\frac{r_{0}}{4R}\right)^4$, we
get \eqref{gaugeisotr}.

\section*{Appendix~C: Calculation of the Hamiltonian}
\setcounter{equation}{0}
\renewcommand{\theequation}{C\arabic{equation}}
\begingroup
\allowdisplaybreaks The result of substitution of \eqref{decomp}
into \eqref{H} can be represented as the sum of three terms
(corresponding to $k=r,\theta,\varphi$ in \eqref{H} respectively)
\begin{equation}
H=H_{1}+H_{2}+H_{3},
\end{equation}
where
\begin{align}\nonumber
H_{1}&=\frac{1}{2}\sum\limits_{p=1}^{2}\sum\limits_{p'=1}^{2}\sum\limits_{j=1}^{\infty}\sum
\limits_{m=-j}^{j}\int\limits_{r_{0}}^{\infty}\frac{dr}{r^{2}}\int\frac{dE}{\sqrt{2E}}
\int\frac{dE'}{\sqrt{2E'}}\,c_{jp}(E)c_{jp'}(E')F_{jp}\left(E,r\right)F_{jp'}\left(E',r\right)\\\nonumber
&\times\left[\left({E'}^{2}-EE'\right)\left(e^{-i(E+E')t}a_{jmp}(E)a_{j(-m)p'}(E')+\textrm{h.c.}\right)\right.\\\label{H1}
&\left.+\left({E'}^{2}+EE'\right)\left(e^{i(E-E')t}a_{jmp}^{\dag}(E)a_{jmp'}(E')+\textrm{h.c.}\right)\right],
\end{align}
integration with respect to $\theta$, $\varphi$ was performed, and
orthogonality conditions for spherical harmonics were used;
\begin{align}\nonumber
H_{2}&=\frac{1}{2}\sum\limits_{p=1}^{2}\sum\limits_{p'=1}^{2}\sum\limits_{j=1}^{\infty}\sum\limits_{j'=1}^{\infty}\sum
\limits_{m=-j}^{j}\sum\limits_{m'=-j'}^{j'}\int
d\varphi\int\sin\theta
d\theta\int\limits_{r_{0}}^{\infty}dr\frac{r}{r-r_{0}}\int\frac{dE}{\sqrt{2E}}
\int\frac{dE'}{\sqrt{2E'}}\\\nonumber
&\times\left[\left({E'}^{2}-EE'\right)\left(e^{-i(E+E')t}C_{2,jmp}(E,r,\theta,\varphi)C_{2,j'm'p'}(E',r,\theta,\varphi)+\textrm{h.c.}\right)\right.\\\label{H2}
&\left.+\left({E'}^{2}+EE'\right)\left(e^{i(E-E')t}C_{2,jmp}^{\dag}(E,r,\theta,\varphi)C_{2,j'm'p'}(E',r,\theta,\varphi)+\textrm{h.c.}\right)\right]
\end{align}
with
\begin{align}\nonumber
C_{2,jmp}(E)&=\frac{1}{j(j+1)}\frac{r-r_{0}}{r}\,c_{jp}(E)\partial_{r}F_{jp}\left(E,r\right)\partial_{\theta}Y_{jm}(\theta,\varphi)a_{jmp}(E)\\
&+\frac{i}{\sqrt{j(j+1)}\sin\theta}\,F_{jp}\left(E,r\right)\partial_{\varphi}Y_{jm}(\theta,\varphi)b_{jmp}(E);
\end{align}
and
\begin{align}\nonumber
H_{3}&=\frac{1}{2}\sum\limits_{p=1}^{2}\sum\limits_{p'=1}^{2}\sum\limits_{j=1}^{\infty}\sum\limits_{j'=1}^{\infty}\sum
\limits_{m=-j}^{j}\sum\limits_{m'=-j'}^{j'}\int
d\varphi\int\frac{d\theta}{\sin\theta}\int\limits_{r_{0}}^{\infty}dr\frac{r}{r-r_{0}}\int\frac{dE}{\sqrt{2E}}
\int\frac{dE'}{\sqrt{2E'}}\\\nonumber
&\times\left[\left({E'}^{2}-EE'\right)\left(e^{-i(E+E')t}C_{3,jmp}(E,r,\theta,\varphi)C_{3,j'm'p'}(E',r,\theta,\varphi)+\textrm{h.c.}\right)\right.\\\label{H3}
&\left.+\left({E'}^{2}+EE'\right)\left(e^{i(E-E')t}C_{3,jmp}^{\dag}(E,r,\theta,\varphi)C_{3,j'm'p'}(E',r,\theta,\varphi)+\textrm{h.c.}\right)\right]
\end{align}
with
\begin{align}\nonumber
C_{3,jmp}(E)&=\frac{1}{j(j+1)}\frac{r-r_{0}}{r}\,c_{jp}(E)\partial_{r}F_{jp}\left(E,r\right)\partial_{\varphi}Y_{jm}(\theta,\varphi)a_{jmp}(E)\\
&-\frac{i\sin\theta}{\sqrt{j(j+1)}}\,F_{jp}\left(E,r\right)\partial_{\theta}Y_{jm}(\theta,\varphi)b_{jmp}(E).
\end{align}
In order to slightly simplify the presentation of the results, in
the formulas presented above the normal ordering of the operators
$a_{jmp}(E)$, $a^{\dagger}_{jmp}(E)$, $b_{jmp}(E)$ and
$b^{\dagger}_{jmp}(E)$ is used and infinite $c$-number terms are
skipped.

First, let us consider the terms that are linear in $c_{jp}(E)$.
The corresponding terms come from \eqref{H2} and \eqref{H3} and
take the form
\begin{align}\nonumber
&\sin\theta\,C_{2,jmp}(E,r,\theta,\varphi)C_{2,j'm'p'}(E',r,\theta,\varphi)
+\frac{1}{\sin\theta}\,C_{3,jmp}(E,r,\theta,\varphi)C_{3,j'm'p'}(E',r,\theta,\varphi)\\\nonumber
&\to\frac{ic_{jp}(E)}{j(j+1)\sqrt{j'(j'+1)}}\frac{r-r_{0}}{r}\,\partial_{r}F_{jp}\left(E,r\right)F_{j'p'}\left(E',r\right)a_{jmp}(E)b_{j'm'p'}(E')\\
&\times\Bigl(\partial_{\theta}Y_{jm}(\theta,\varphi)\partial_{\varphi}Y_{j'm'}(\theta,\varphi)
-\partial_{\varphi}Y_{jm}(\theta,\varphi)\partial_{\theta}Y_{j'm'}(\theta,\varphi)\Bigr)
+[j,m,p,E\leftrightarrow j',m',p',E'], \label{H23cjp}
\end{align}
the hermitian conjugate of \eqref{H23cjp}, and
\begin{align}\nonumber
&\sin\theta\,C_{2,j'm'p'}^{\dagger}(E',r,\theta,\varphi)C_{2,jmp}(E,r,\theta,\varphi)
+\frac{1}{\sin\theta}\,C_{3,j'm'p'}^{\dagger}(E',r,\theta,\varphi)C_{3,jmp}(E,r,\theta,\varphi)\\\nonumber
&\to\frac{-ic_{jp}(E)}{j(j+1)\sqrt{j'(j'+1)}}\frac{r-r_{0}}{r}\,\partial_{r}F_{jp}\left(E,r\right)F_{j'p'}\left(E',r\right)b_{j'm'p'}^{\dagger}(E')a_{jmp}(E)\\
&\times\Bigl(\partial_{\theta}Y_{jm}(\theta,\varphi)\partial_{\varphi}Y^{*}_{j'm'}(\theta,\varphi)
-\partial_{\varphi}Y_{jm}(\theta,\varphi)\partial_{\theta}Y^{*}_{j'm'}(\theta,\varphi)\Bigr)
+[j,m,p,E\leftrightarrow j',m',p',E']^{\dagger} \label{H23cjpdag}
\end{align}
and its hermitian conjugate. Let us take the term
\begin{equation}\label{H23cjp2}
\sum\limits_{m=-j}^{j}\sum\limits_{m'=-j'}^{j'}\int d\varphi\int
d\theta
\Bigl(\partial_{\theta}Y_{jm}(\theta,\varphi)\partial_{\varphi}Y_{j'm'}(\theta,\varphi)
-\partial_{\varphi}Y_{jm}(\theta,\varphi)\partial_{\theta}Y_{j'm'}(\theta,\varphi)\Bigr),
\end{equation}
which follows from \eqref{H2}, \eqref{H3} and \eqref{H23cjp}. It
is clear that \eqref{H23cjp2} is equal to
\begin{align}\nonumber
&\sum\limits_{m=-j}^{j}\sum\limits_{m'=-j'}^{j'}\int d\varphi\int
d\theta
\Bigl(\partial_{\theta}Y_{jm}(\theta,\varphi)\partial_{\varphi}Y_{j'm'}(\theta,\varphi)
+Y_{jm}(\theta,\varphi)\partial_{\theta}\partial_{\varphi}Y_{j'm'}(\theta,\varphi)\Bigr)\\
&=\sum\limits_{m=-j}^{j}\sum\limits_{m'=-j'}^{j'}\int d\varphi\,
Y_{jm}(\theta,\varphi)\partial_{\varphi}Y_{j'm'}(\theta,\varphi)\Bigl|_{0}^{\pi}
=i\sum\limits_{m=-\textrm{min}\{j,j'\}}^{\textrm{min}\{j,j'\}}m
Y_{jm}(\theta,0)Y_{j'(-m)}(\theta,0)\Bigl|_{0}^{\pi}. \label{H23cjp3}
\end{align}
Since $Y_{jm}(\theta,0)\sim P_{j}^{m}(\cos\theta)$, where
$P_{j}^{m}(\cos\theta)$ are the associated Legendre polynomials,
whereas $P_{j}^{m}(\pm 1)=0$ for $m\neq 0$, we get
\begin{equation}
\sum\limits_{m=-\textrm{min}\{j,j'\}}^{\textrm{min}\{j,j'\}}m
Y_{jm}(\theta,0)Y_{j'(-m)}(\theta,0)\Bigl|_{0}^{\pi}=0.
\end{equation}

A fully analogous procedure can be performed for the term
\begin{equation}\label{H23cjp2dag}
\sum\limits_{m=-j}^{j}\sum\limits_{m'=-j'}^{j'}\int d\varphi\int
d\theta
\Bigl(\partial_{\theta}Y_{jm}(\theta,\varphi)\partial_{\varphi}Y^{*}_{j'm'}(\theta,\varphi)
-\partial_{\varphi}Y_{jm}(\theta,\varphi)\partial_{\theta}Y^{*}_{j'm'}(\theta,\varphi)\Bigr),
\end{equation}
which follows from \eqref{H2}, \eqref{H3} and \eqref{H23cjpdag},
resulting in
\begin{equation}
\sum\limits_{m=-\textrm{min}\{j,j'\}}^{\textrm{min}\{j,j'\}}m
Y_{jm}(\theta,0)Y^{*}_{j'm}(\theta,0)\Bigl|_{0}^{\pi}=0.
\end{equation}
Thus, the terms in $H$ that are linear in $c_{jp}(E)$ vanish.

Next, let us consider the rest of the terms proportional to
${E'}^{2}-EE'$. First we take the terms that come from $H_{2}$
(formula \eqref{H2}) and $H_{3}$ (formula \eqref{H3}):
\begin{align}\nonumber
&\frac{1}{2}\sum\limits_{p=1}^{2}\sum\limits_{p'=1}^{2}\sum\limits_{j=1}^{\infty}\sum\limits_{j'=1}^{\infty}\sum
\limits_{m=-j}^{j}\sum\limits_{m'=-j'}^{j'}\int d\varphi\int
d\theta\int\limits_{r_{0}}^{\infty}dr\frac{r}{r-r_{0}}\int\frac{dE}{\sqrt{2E}}
\int\frac{dE'}{\sqrt{2E'}}\left({E'}^{2}-EE'\right)
\\\nonumber
&\times\biggl[ e^{-i(E+E')t} \biggl(\frac{(r-r_{0})^{2}c_{jp}(E)c_{j'p'}(E')}{r^{2}j(j+1)j'(j'+1)}\,
\partial_{r}F_{jp}\left(E,r\right)\partial_{r}F_{j'p'}\left(E',r\right)a_{jmp}(E)a_{j'm'p'}(E')
\\\nonumber&-
\frac{1}{\sqrt{j(j+1)j'(j'+1)}}\,F_{jp}\left(E,r\right)F_{j'p'}\left(E',r\right)b_{jmp}(E)b_{j'm'p'}(E')\biggr) \\\label{H23aux1}&\times\sin\theta
\left(-\frac{1}{\sin\theta}\partial_{\theta}\Bigl(\sin\theta\partial_{\theta}Y_{jm}(\theta,\varphi)\Bigr)-
\frac{1}{\sin^{2}\theta}\partial^{2}_{\varphi}Y_{jm}(\theta,\varphi)\right)Y_{j'm'}(\theta,\varphi)+\textrm{h.c.}\biggr],
\end{align}
where integration by parts with respect to  $\varphi$ and $\theta$
is performed  (here the corresponding surface terms at $\theta=0$
and $\theta=\pi$ are standard, they do not need special treatment
like those in \eqref{H23cjp3} and vanish automatically because of
the extra $\sin\theta$ that was absent in the surface terms in
\eqref{H23cjp3}). Using \eqref{Y_eq}, formula \eqref{H23aux1} can
be rewritten as
\begin{align}\nonumber
&\frac{1}{2}\sum\limits_{p=1}^{2}\sum\limits_{p'=1}^{2}\sum\limits_{j=1}^{\infty}\sum\limits_{j'=1}^{\infty}\sum
\limits_{m=-j}^{j}\sum\limits_{m'=-j'}^{j'}\int d\varphi\int
d\theta\int\limits_{r_{0}}^{\infty}dr\frac{r}{r-r_{0}}\int\frac{dE}{\sqrt{2E}}
\int\frac{dE'}{\sqrt{2E'}}\left({E'}^{2}-EE'\right)
\\\nonumber
&\times\biggl[ e^{-i(E+E')t} \biggl(\frac{(r-r_{0})^{2}c_{jp}(E)c_{j'p'}(E')}{r^{2}j'(j'+1)}\,
\partial_{r}F_{jp}\left(E,r\right)\partial_{r}F_{j'p'}\left(E',r\right)a_{jmp}(E)a_{j'm'p'}(E')
\\\nonumber
&-\frac{\sqrt{j(j+1)}}{\sqrt{j'(j'+1)}}\,F_{jp}\left(E,r\right)F_{j'p'}\left(E',r\right)b_{jmp}(E)b_{j'm'p'}(E')\biggr) \sin\theta
Y_{jm}(\theta,\varphi)Y_{j'm'}(\theta,\varphi)+\textrm{h.c.}\biggr]
\\\nonumber
&=
\frac{1}{2}\sum\limits_{p=1}^{2}\sum\limits_{p'=1}^{2}\sum\limits_{j=1}^{\infty}\sum
\limits_{m=-j}^{j}\int\frac{dE}{\sqrt{2E}}
\int\frac{dE'}{\sqrt{2E'}}\int\limits_{r_{0}}^{\infty}dr\left({E'}^{2}-EE'\right)
\\\nonumber
&\times \biggl[e^{-i(E+E')t} \left(-\frac{c_{jp}(E)c_{jp'}(E')}{j(j+1)}\,
\partial_{r}\left\{\frac{r-r_{0}}{r}\partial_{r}F_{jp}\left(E,r\right)\right\}F_{jp'}\left(E',r\right)a_{jmp}(E)a_{j(-m)p'}(E')\right.
\\\label{H23aux2}&-\left.
\frac{r}{r-r_{0}}\,F_{jp}\left(E,r\right)F_{jp'}\left(E',r\right)b_{jmp}(E)b_{j(-m)p'}(E')\right)+\textrm{h.c.}\biggr],
\end{align}
where the orthogonality conditions for the spherical harmonics
$Y_{jm}(\theta,\varphi)$ were used and integration by parts with
respect to $r$ was performed. Combining \eqref{H23aux2} with the
term proportional to ${E'}^{2}-EE'$ in $H_{1}$ (formula
\eqref{H1}), and using Eq.~\eqref{radial_eq_4} and orthogonality
condition \eqref{norm}, we get
\begin{align}\nonumber
&\frac{1}{2}\sum\limits_{p=1}^{2}\sum\limits_{p'=1}^{2}\sum\limits_{j=1}^{\infty}\sum
\limits_{m=-j}^{j}\int\frac{dE}{\sqrt{2E}}
\int\frac{dE'}{\sqrt{2E'}}\int\limits_{r_{0}}^{\infty}dr\left({E'}^{2}-EE'\right)
\\\nonumber
&\times \biggl[e^{-i(E+E')t} \biggl(-\frac{c_{jp}(E)c_{jp'}(E')}{j(j+1)}\,
\left\{\partial_{r}\left(\frac{r-r_{0}}{r}\partial_{r}F_{jp}\left(E,r\right)\right)
-\frac{j(j+1)}{r^{2}}F_{jp}\left(E,r\right)\right\} \\\nonumber & \times
F_{jp'}\left(E',r\right) a_{jmp}(E)a_{j(-m)p'}(E') -
\frac{r}{r-r_{0}}\,F_{jp}\left(E,r\right)F_{jp'}\left(E',r\right)b_{jmp}(E)b_{j(-m)p'}(E')\biggr)+\textrm{h.c.}\biggr]\\\nonumber
&=\frac{1}{2}\sum\limits_{p=1}^{2}\sum\limits_{p'=1}^{2}\sum\limits_{j=1}^{\infty}\sum
\limits_{m=-j}^{j}\int\frac{dE}{\sqrt{2E}}
\int\frac{dE'}{\sqrt{2E'}}\left({E'}^{2}-EE'\right)
\\\nonumber
&\times\biggl[e^{-i(E+E')t}\biggl(\left\{\frac{c_{jp}(E)c_{jp'}(E')E^{2}}{j(j+1)}\,a_{jmp}(E)a_{j(-m)p'}(E')-b_{jmp}(E)b_{j(-m)p'}(E')\right\}
\\\nonumber&\times\int\limits_{r_{0}}^{\infty}dr
\frac{r}{r-r_{0}}\,F_{jp}\left(E,r\right)F_{jp'}\left(E',r\right)\biggr)+\textrm{h.c.}\biggr]\\\nonumber
&=\frac{1}{2}\sum\limits_{p=1}^{2}\sum\limits_{j=1}^{\infty}\sum
\limits_{m=-j}^{j}\int\frac{dE}{2E}\left({E}^{2}-E^{2}\right)\\
&\times\left[e^{-i2Et}\left(\frac{c_{jp}^{2}(E)E^{2}}{j(j+1)}\,a_{jmp}(E)a_{j(-m)p}(E)-b_{jmp}(E)b_{j(-m)p}(E)\right)+\textrm{h.c.}\right]=0.
\end{align}
Thus, the terms in $H$ that are proportional to ${E'}^{2}-EE'$
also vanish.

A fully analogous procedure can be performed for the terms in $H$
that are proportional to ${E'}^{2}+EE'$, resulting in
\begin{align}\nonumber
&\frac{1}{2}\sum\limits_{p=1}^{2}\sum\limits_{p'=1}^{2}\sum\limits_{j=1}^{\infty}\sum
\limits_{m=-j}^{j}\int\frac{dE}{\sqrt{2E}}
\int\frac{dE'}{\sqrt{2E'}}\int\limits_{r_{0}}^{\infty}dr\left({E'}^{2}+EE'\right) \\\nonumber
&\times \biggl[e^{i(E-E')t} \biggl(-\frac{c_{jp}(E)c_{jp'}(E')}{j(j+1)}\,
\left\{\partial_{r}\left(\frac{r-r_{0}}{r}\partial_{r}F_{jp}\left(E,r\right)\right)
-\frac{j(j+1)}{r^{2}}F_{jp}\left(E,r\right)\right\} \\\nonumber &\times
F_{jp'}\left(E',r\right) a^{\dagger}_{jmp}(E)a_{jmp'}(E')
+\frac{r}{r-r_{0}}\,F_{jp}\left(E,r\right)F_{jp'}\left(E',r\right)b^{\dagger}_{jmp}(E)b_{jmp'}(E')\biggr)+\textrm{h.c.}\biggr]\\\nonumber
&=\frac{1}{2}\sum\limits_{p=1}^{2}\sum\limits_{p'=1}^{2}\sum\limits_{j=1}^{\infty}\sum
\limits_{m=-j}^{j}\int\frac{dE}{\sqrt{2E}}
\int\frac{dE'}{\sqrt{2E'}}\left({E'}^{2}+EE'\right)
\\\nonumber
&\times\biggl[e^{i(E-E')t}\biggl(\left\{\frac{c_{jp}(E)c_{jp'}(E')E^{2}}{j(j+1)}\,a^{\dagger}_{jmp}(E)a_{jmp'}(E')+b^{\dagger}_{jmp}(E)b_{jmp'}(E')\right\}
\\\nonumber&\times\int\limits_{r_{0}}^{\infty}dr
\frac{r}{r-r_{0}}\,F_{jp}\left(E,r\right)F_{jp'}\left(E',r\right)\biggr)+\textrm{h.c.}\biggr]\\\nonumber
&=\frac{1}{2}\sum\limits_{p=1}^{2}\sum\limits_{j=1}^{\infty}\sum
\limits_{m=-j}^{j}\int dE\,E
\left(\frac{c^{2}_{jp}(E)E^{2}}{j(j+1)}\,a^{\dagger}_{jmp}(E)a_{jmp}(E)+b^{\dagger}_{jmp}(E)b_{jmp}(E)\right)+\textrm{h.c.}\\
&=\sum\limits_{p=1}^{2}\sum\limits_{j=1}^{\infty}\sum
\limits_{m=-j}^{j}\int dE\,E
\left(\frac{c^{2}_{jp}(E)E^{2}}{j(j+1)}\,a^{\dagger}_{jmp}(E)a_{jmp}(E)+b^{\dagger}_{jmp}(E)b_{jmp}(E)\right).
\end{align}
One can see that the coefficient in front of
$b^{\dagger}_{jmp}(E)b_{jmp}(E)$ is correct, which confirms the
validity of  normalization condition \eqref{norm}. In order to get
the correct coefficient in front of
$a^{\dagger}_{jmp}(E)a_{jmp}(E)$, one should set
\begin{equation}
c_{jp}(E)=\frac{\sqrt{j(j+1)}}{E}.
\end{equation}
Finally, we get \eqref{H-final}.

\end{document}